# Investigation of superconducting gap of high-entropy telluride AgInSnPbBiTe$_5$

Asato Seshita[1], Hirotaka Okabe[2,3], Riad Kasem[1], Yuto Watanabe[1], Jumpei G. Nakamura[2,4], Shoichiro Nishimura[2,4], Kensei Terashima[5], Ryo Matsumoto[5], Yoshihiko Takano[5], Aichi Yamashita[1], Masaki Fujita[3], Yoshikazu Mizuguchi[1]*

1 Department of Physics, Tokyo Metropolitan University, Hachioji 192-0397, Japan
2 Muon Science Division, Institute of Materials Structure Science, High Energy Accelerator Research Organization (KEK), Tsukuba 305-0801, Japan
3 Institute for Materials Research, Tohoku University, Sendai 980-8577, Japan
4 Materials and Life Science Division, J-PARC Center, Tokai 319-1106, Japan
5 National Institute for Materials Science, Tsukuba 305-0047, Japan

*mizugu@tmu.ac.jp

Abstract

We performed transverse-field muon spin relaxation/rotation (TF-$\mu$SR) on a high-entropy-type (HE-type) superconductor AgInSnPbBiTe$_5$. The emergence of bulk superconducting states was confirmed from magnetic susceptibility, specific heat, and $\mu$SR. The superconducting gap 2$\Delta$(0) estimated from $\mu$SR was clearly larger than that expected from conventional weak-coupling phonon-mediated model, suggesting the strong-coupling nature of superconductivity. In addition, a long penetration depth of 3.21(7) μm was obtained. The strong-coupling nature of superconductivity and the long penetration depth are similar to the trends observed in the other HE-type superconductors (HE alloys and transition-metal zirconides), which may be universal feature of HE-type superconductors.

**Keywords:** Superconductivity, superconducting gap, $\mu$SR, metal telluride, high-pressure synthesis

**Statements and Declarations**

The authors declare no competing interests.

**Data availability statement**

All the data presented in the paper can be provided from the corresponding author by a reasonable request.

1. **Introduction**

Recently, high-entropy (HE) materials, which are highly-disordered materials possessing multiple-element solution resulting in high configurational entropy, have been actively studied. The simplest HE materials are HE alloys (HEAs), which alloys typically containing five or more elements solving in a crystallographic site with a concentration of 5–35 at% [1–3]. The above-described criterion for element solution satisfies a high value of configurational entropy of mixing ($\Delta S_{mix}$), exceeding $1.5R$, where $R$ is gas constant. $\Delta S_{mix}$ is defined as $\Delta S_{mix} = -R\Sigma_i c_i \ln c_i$, where $c_i$ is composition of the component $i$. HEAs and similar HE compounds, where one or more crystallographic sites of compounds are substituted by multiple elements, possess merit on the engineering aspects. For example, HEAs exhibit improved mechanical properties and stability or high performance in high temperature and/or extreme conditions [1,2]. Furthermore, superior catalytic and thermoelectric properties have been observed [4–9]. As well as the application merit, the highly disordered local structures and nearly-random bonding should be important issues in the field of materials science. Actually, HE superconductors with various types of crystal structures and constituent elements have been developed [10–23], and some of the HE superconductors exhibit robustness of superconductivity against applied external pressure [24,25]. In Ref. 26, we reported on the glassy atomic vibration and blurry electronic band structure in a AgInSnPbBiTe$_5$ superconductor, and the origin of the unconventional pressure dependence of transition temperature ($T_c$) is discussed with the anomalous glassy phonon and electronic states.

In this paper, we study the superconducting properties of a HE-type metal telluride AgInSnPbBiTe$_5$ with a NaCl-type structure (space group: $Fm$-$3m$, #225) using muon spin relaxation/rotation ($\mu$SR), which is a powerful tool for discussing superconducting gap [27]. As shown in Fig. 1(a), the metal site is in the HEA configuration with the solution of five different metals. From the temperature evolution of magnetic penetration depth ($\lambda$), the superconducting gap and pairing symmetry can be discussed. Several $\mu$SR studies on superconducting properties of HEAs and HE compounds revealed that the HE superconductors [28,29] tend to show a value of $2\Delta(0)/k_B T_c$ greater than 3.53 expected from the Bardeen-Cooper-Schrieffer (BCS) phonon-mediated weak-coupling model; $\Delta(0)$ and $k_B$ are a superconducting gap at 0 K and the Boltzmann constant, respectively [30]. The examples of $2\Delta(0)/k_B T_c$ estimated using $\mu$SR are summarized in Fig. S1. Furthermore, there was a clear trend that the increase in $\Delta S_{mix}$ results in an increase in $2\Delta(0)/k_B T_c$ for transition-metal ($Tr$) zirconides $Tr$Zr$_2$ [29], and the $T_c$ of HE-type samples deviates from the conventional line in the Uemura plot [28,29]. If another example of strong (or moderate) coupling characteristics of superconductivity in HE materials, universal features of HE superconductors would be obtained, and that will be useful for clarifying the essential effects of HE-alloying on superconducting states. Here, we show the strong-coupling nature of the AgInSnPbBiTe$_5$ superconductor.

2. **Experimental methods**

Polycrystalline powders of AgInSnPbBiTe$_5$ were firstly prepared by solid-state reaction in an evacuated quartz tube. High-pressure annealing of the prepared precursor powders was performed to obtain single-phase polycrystalline samples of AgInSnPbBiTe$_5$ using a cubic-anvil-type press system (CT factory). High-pressure annealing conditions were 3 GPa and 500°C for 30 min. The details of sample preparation are reported in Refs. 20, 23, and 25.

Powder X-ray diffraction (XRD) was performed using a MiniFlex600 (RIGAKU) diffractometer with Cu Kα radiation and a D/teX-Ultra detector by a conventional $\theta$-$2\theta$ method. The crystal-structure parameters were refined using the Rietveld method with RIETAN-FP software [31], and the crystal structure was visualized using VESTA software [32]. The chemical composition of the selected sample was examined by energy-dispersive X-ray spectroscopy (EDX) on a scanning electron microscope TM-3030 (Hitachi Hightech) equipped with an EDX-SwiftED analyzer (Oxford). The EDX analysis was performed at five different points. To investigate the superconducting properties of AgInSnPbBiTe$_5$, the temperature dependence of magnetic susceptibility ($4\pi\chi$) was measured using a superconducting quantum interference device (SQUID) with a Magnetic Property Measurement System (MPMS, Quantum Design) after both zero-field cooling (ZFC) and field cooling (FC) occurred.

Specific heat ($C$) measurements were performed using the thermal relaxation method with a Physical Property Measurement System (PPMS Dynacool, Quantum Design). The sample was fixed on a sample stage with the Apiezon N grease.

Transverse field (TF) $\mu$SR measurements were performed using an ARTEMIS spectrometer installed on S1 area in MLF, J-PARC (proposal No.: 2024A0076). The measurements were performed under a TF of $H$ = 246 Oe using a $^3$He cryostat.

## 3. Results and discussion

### 3.1 Sample characterization

Powder XRD pattern for AgInSnPbBiTe$_5$ is shown in Fig. 1(b). Although the tiny diffraction peak of BN ($R3m$, #160), which was used in the high-pressure annealing, was detected (see Fig. S1 for the Rietveld fitting), the other peaks were assigned to the NaCl-type phase with no peak splitting, indicating that homogeneous AgInSnPbBiTe$_5$ was successfully synthesized. The lattice constant is estimated as 6.25405(5) Å. No compositional segregation was detected by EDX mapping (Fig. 1(c)). The average chemical composition of the obtained sample is estimated as Ag$_{1.01}$In$_{1.00}$Sn$_{1.00}$Pb$_{0.88}$Bi$_{1.14}$Te$_{4.96}$. The estimated $\Delta S_{mix}$ for the M site using average chemical composition is 1.61$R$.

### 3.2 Magnetic susceptibility and specific heat

Figure 2(a) shows the temperature dependence of $4\pi\chi$ of AgInSnPbBiTe$_5$. The $4\pi\chi$ data is corrected by assuming the demagnetization effect. Large diamagnetic signals corresponding to the emergence of superconducting states are observed below 2.5 K, and the estimated $T_c$ was 2.44 K; $T_c$ was estimated as the crossing point of two lines as shown in the inset of Fig. 2(a).

Figure 2(b) shows the temperature dependences of $C$ under magnetic fields of $\mu_0H$ = 0.0−1.0 T. Clear jumps were observed at $\mu_0H \leq 0.6$ T, suggesting the emergence of bulk superconductivity. As shown in Fig. 2(c), the estimated $T_c$s were plotted, and the upper critical field at 0 K ($\mu_0H_{c2}(0)$) was estimated as 0.74 T by assuming the Werthamer-Helfand-Hohenberg model [33]. The low-temperature specific heat can be described as $C/T = \gamma + \beta T^2 + \delta T^4$, where $\gamma$, $\beta$, and $\delta$ represent contributions of electron, lattice, and anharmonicity term of the lattice, respectively. The values of $\gamma$, $\beta$, and $\delta$ were estimated as 2.6(2) mJ mol$^{-1}$ K$^{-2}$, 0.70(6) mJ mol$^{-1}$ K$^{-4}$, and 0.043(5) mJ mol$^{-1}$ K$^{-6}$, respectively, from fitting the data for $\mu_0H$ = 1.0 T. The Debye temperature $\Theta_D$ of 177 K was calculated from $\beta = 12\pi^4 NR/5\Theta_D^3$, where $N$ is the number of atoms in the formula unit. To examine the magnitude of electronic specific heat jump ($\Delta C_{ele}$), the temperature dependence of $C_{ele}/T$ is plotted by subtracting the lattice contribution ($\beta T^2 + \delta T^4$) as shown in Fig. 2(d). $T_c$ and $\Delta C_{ele}$ at $T_c$ were estimated by considering the entropy balance, and the estimated values are $T_c$ = 2.44 and $\Delta C_{ele}$ = 1.14$\gamma T_c$. The $\Delta C_{ele}$ of AgInSnPbBiTe$_5$ is slightly smaller than 1.43$\gamma T_c$, which is expected from the BCS theory with a weak-coupling superconductor, but the comparable $\Delta C_{ele}$ proves the emergence of bulk superconductivity.

3.3 TF-$\mu$SR

To obtain information on the superconducting gap information, TF-$\mu$SR measurement has been performed. Magnetic field of $H$ = 246 Oe, which is in between the lower critical field at 0 K($\mu_0H_{c1}(0)$) and the $\mu_0H_{c2}(0)$, was applied above the $T_c$, followed by sample cooling to 0.69 K. Figure 3 shows the TF-$\mu$SR asymmetry spectrum at 0.69 K for AgInSnPbBiTe$_5$. The spectrum exhibits damping oscillation, indicating that the sample falls into the flux-line-lattice state to exert strongly inhomogeneous internal field to implanted muons. The TF-$\mu$SR signal is best fitted with the following model function [28]:

$$A(t) = A\exp\left(-\frac{1}{2}\sigma^2 t^2\right)\cos(\gamma_\mu Bt + \phi) + \sum_{n=1}^{2} A_{BGn}\exp\left(-\frac{1}{2}\sigma_{BGn}^2 t^2\right)\cos(\gamma_\mu B_{BGn}t + \phi),$$

where $A$ and $A_{BGn}$ are the asymmetry contributions from sample and sample holder, $\sigma$ and $\sigma_{BGn}$ are relaxation rate contributions from sample and sample holder, $\gamma_\mu$ is muon gyromagnetic ratio, $B$ and $B_{BGn}$ are mean field contributions from sample and sample holder, and $\phi$ is the initial phase offset. To accurately examine the small $\sigma$, fitting was performed with long time range spectrum (up to 15 μs). The $\sigma$ includes both the temperature-independent depolarization $\sigma_N$, which comes from the static field arising due to the nuclear magnetic moment, and the contribution of the field variation from the flux-line-lattice, given as $\sigma^2 = \sigma_N^2 + \sigma_{FLL}^2$. For a triangular lattice, the temperature dependence of the magnetic penetration depth $\lambda(T)$ can be expressed by $\frac{\sigma_{FLL}(T)^2}{\gamma_\mu^2} = \frac{0.00371\phi_0^2}{\lambda^4(T)}$, where $\gamma_\mu/2\pi$ = 135.5 MHz/T is the muon gyromagnetic ratio, and $\phi_0$ is a flux quantum. The temperature dependence of $\lambda^{-2}$ is shown in Fig. 4. To examine the $\lambda(0)$ and $\Delta(0)$, the data was fitted with a superconducting gap function. Assuming an isotropic $s$-wave gap, the temperature dependent relation between the $\lambda(0)$ and $\Delta(0)$ can be described as

$$\frac{\lambda(0)^2}{\lambda(T)^2} = \frac{\Delta(T)}{\Delta(0)} \tanh\left[\frac{\Delta(T)}{2k_B T}\right],$$

which is relation derived within the BCS scheme, i.e. $2\Delta(0) = 3.53 k_B T_c$ and in the dirty limit [29]. Here, we obtained $T_c$ of 1.88 K from the fitting. The lower $T_c$ than that shown in Fig. 2 is due to the strain-release effect and applied magnetic field. In AgInSnPbBiTe$_5$, the flesh sample shows $T_c$ close to 2.5 K as shown in Fig. 2, but after several weeks, the high-pressure-annealing strain is released, and $T_c$ decreases to about 2 K (See Fig. S2 for the susceptibility data taken after 3 months from the high-pressure synthesis). The current $\mu$SR measurements were performed after 16 days from the high-pressure annealing. The solid line in Fig. 4 is the fit, and the long $\lambda(0)$ of 3.21(7) μm and large $2\Delta(0)/k_B T_c$ of 10(2) are obtained. This large $2\Delta(0)$ clearly larger than the BCS value suggests that the AgInSnPbBiTe$_5$ is possibly possessing strong-coupling nature of superconductivity. The examples of large $2\Delta/k_B T_c$ in Fe- and Cr-based samples are summarized in Table S1. The estimated $2\Delta/k_B T_c$ in the current sample is close to the largest value in the Fe-based superconductors. The large penetration depth would be caused by highly inhomogeneous atomic alignments in HE-type material, and such inhomogeneity would influence the homogeneous formation of the superconducting gap, which has been observed in other HE-type superconductors examined by $\mu$SR [29]. Therefore, the trend that HE-type superconductors have a strong-coupling nature would be the universal features when the HE configuration of the atoms largely affect the electronic and phonon characteristics and superconducting gap characterization. To confirm the universality, further experimental and theoretical studies are needed. In particular, for AgInSnPbBiTe$_5$, low-temperature $C$ measurements using a $^3$He or dilution system will be needed to examine superconducting gap.

## 4. Summary

We synthesized polycrystalline samples of AgInSnPbBiTe$_5$, which is a HE-type superconductor, using high-pressure annealing. From the powder XRD and EDX mapping, homogeneous single-phase quality of the obtained samples was confirmed. Bulk nature of superconductivity was confirmed using magnetic susceptibility and specific heat measurements. TF-$\mu$SR was performed on AgInSnPbBiTe$_5$ under TF of $H$ = 246 Oe. The emergence of bulk superconducting states was confirmed from $\mu$SR, and the estimated $2\Delta(0)/k_B T_c$ of 10 suggested the strong-coupling nature of superconductivity. The trend that the HE-type samples exhibit strong-coupling nature of superconductivity would be a universal feature, which will create new research field of physics for highly disordered superconductors.

**Acknowledgements**

The authors thank E. Kenny, Y. Ikeda, and A. Bhattacharyya for discussion on the results. This project was partly supported by TMU Research Project for Emergent Future Society.

**Figures**

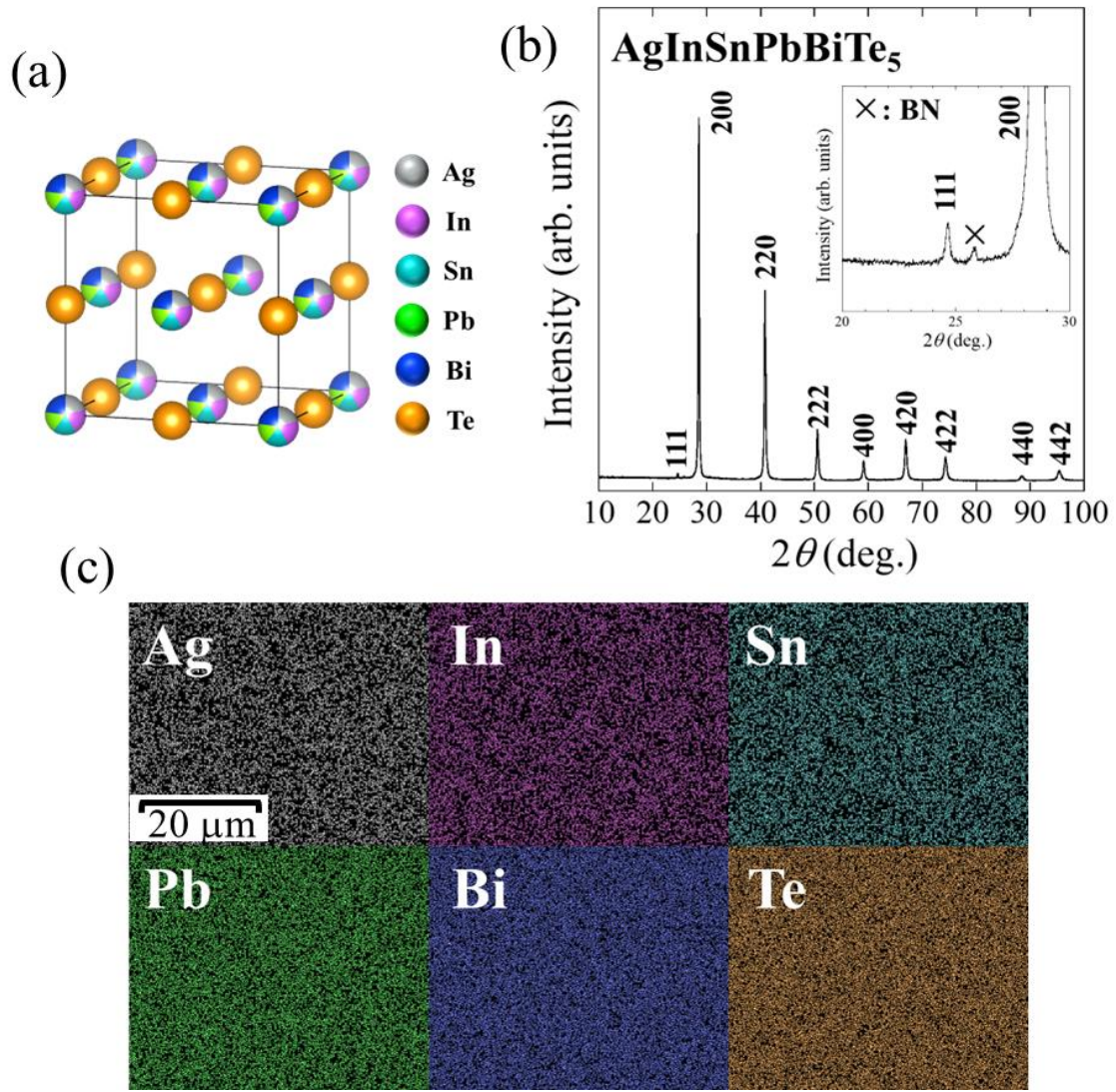

Fig. 1. (a) Schematic image of the crystal structure of AgInSnPbBiTe$_5$. (b) Powder XRD pattern. The numbers are Miller indices. (c) EDX mapping results.

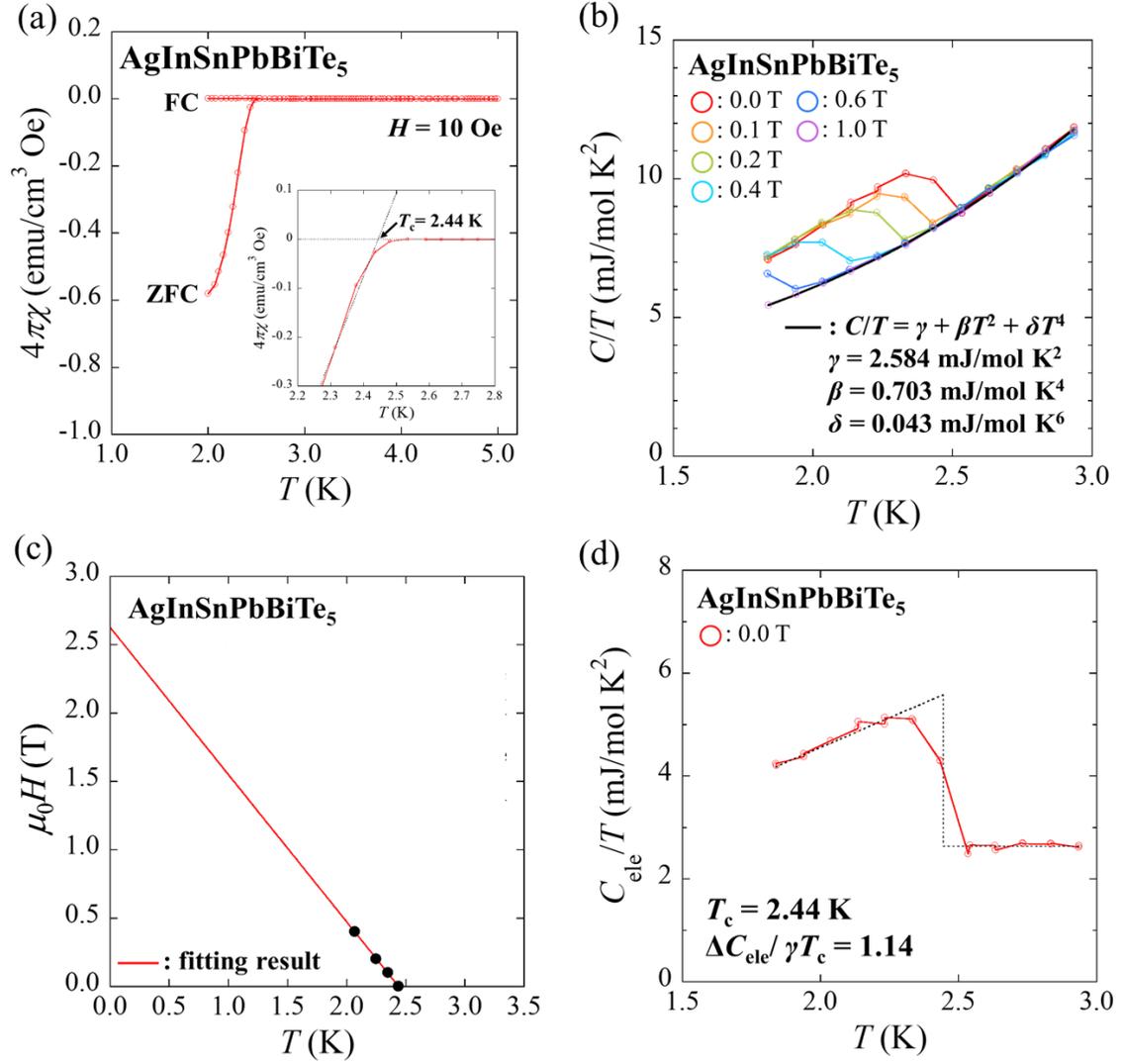

Fig. 2. Temperature dependence of (a) Magnetic susceptibility ($4\pi\chi$) measured at $H$ = 10 Oe and (b) specific heat at $\mu_0 H$ = 0.0−1.0 T for AgInSnPbBiTe$_5$. (c) Magnetic field-temperature phase diagram. (d) Temperature dependence of electronic specific heat $C_{ele}$ in the form of $C_{ele}/T$.

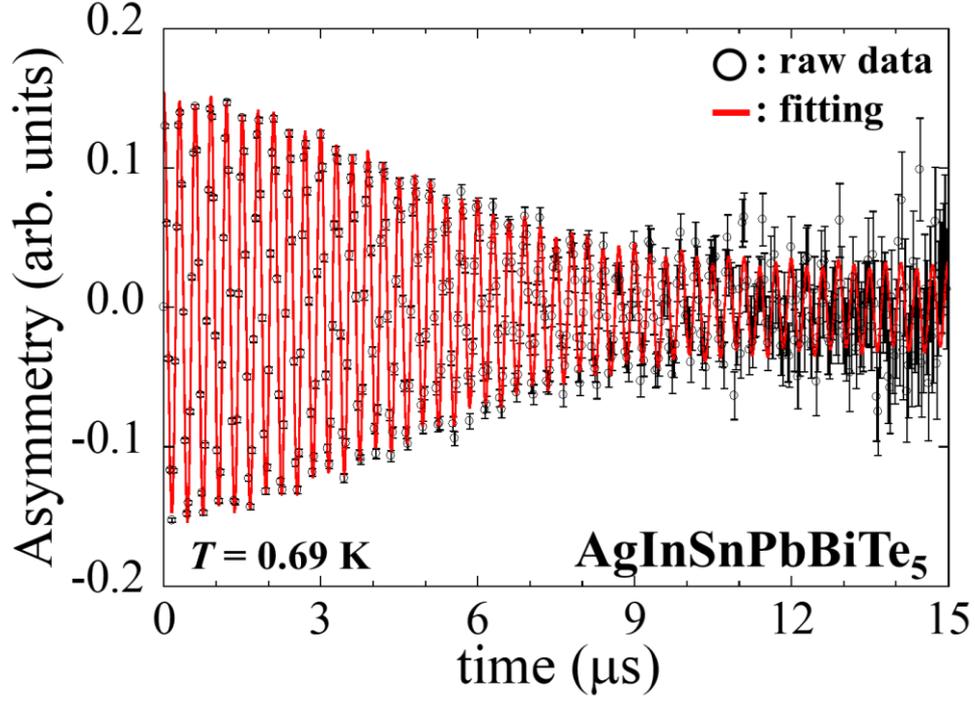

Fig. 3. Obtained asymmetry spectrum at $T$ = 0.32 K and fitting result for AgInSnPbBiTe$_5$.

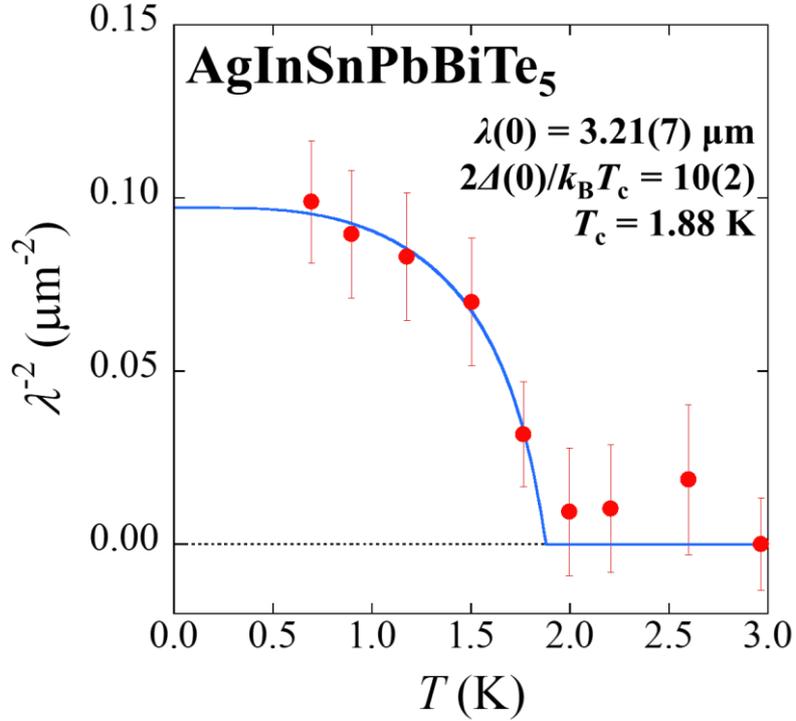

Fig. 4. Temperature dependence of $\lambda^{-2}$ for AgInSnPbBiTe$_5$ and the fitting result with a dirty-limit model.

# Supporting information

# Investigation of superconducting gap of high-entropy telluride AgInSnPbBiTe$_5$

Asato Seshita[1], H. Okabe[2,3], Riad Kasem[1], Y. Watanabe[1], Jumpei G. Nakamura[2,4], Shoichiro Nishimura[2,4], Kensei Terashima[5], Ryo Matsumoto[5], Yoshihiko Takano[5], A. Yamashita[1], Masaki Fujita[3], Yoshikazu Mizuguchi[1]*

1 Department of Physics, Tokyo Metropolitan University, Hachioji 192-0397, Japan
2 Muon Science Division, Institute of Materials Structure Science, High Energy Accelerator Research Organization (KEK), Tsukuba 305-0801, Japan
3 Institute for Materials Research, Tohoku University, Sendai 980-8577, Japan
4 Materials and Life Science Division, J-PARC Center, Tokai 319-1106, Japan
5 National Institute for Materials Science, Tsukuba 305-0047, Japan

*mizugu@tmu.ac.jp

Table S1. Information of $2\Delta(0)/k_BT_c$ estimated using $\mu$SR of various high-entropy superconducting materials and examples of superconductors having a strong-coupling nature. Two values of $2\Delta(0)/k_BT_c$ are obtained from fitting using a multi-gap model.

| Material | $2\Delta(0)/k_BT_c$ | Reference |
|---|---|---|
| AgInSnPbBiTe$_5$ | 10(2) | This study |
| (Fe, Co, Ni, Rh, Ir)Zr$_2$ | 4.69 | [S1] |
| Hf-Nb-Mo-Re-Ru | 3.37 | [S2] |
| Zr-Nb-Mo-Re-Ru | 3.94 | [S2] |
| Nb-Re-Zr-Hf-Ti | 5.31 | [S3] |
| Ba$_{0.6}$K$_{0.4}$Fe$_2$As$_2$ | 7.3, 4.1 | [S4] |
| KCa$_2$Fe$_4$As$_4$F$_2$ | 7.03, 1.28 | [S5] |
| Cs$_2$Cr$_3$As$_2$ | 6 | [S6] |

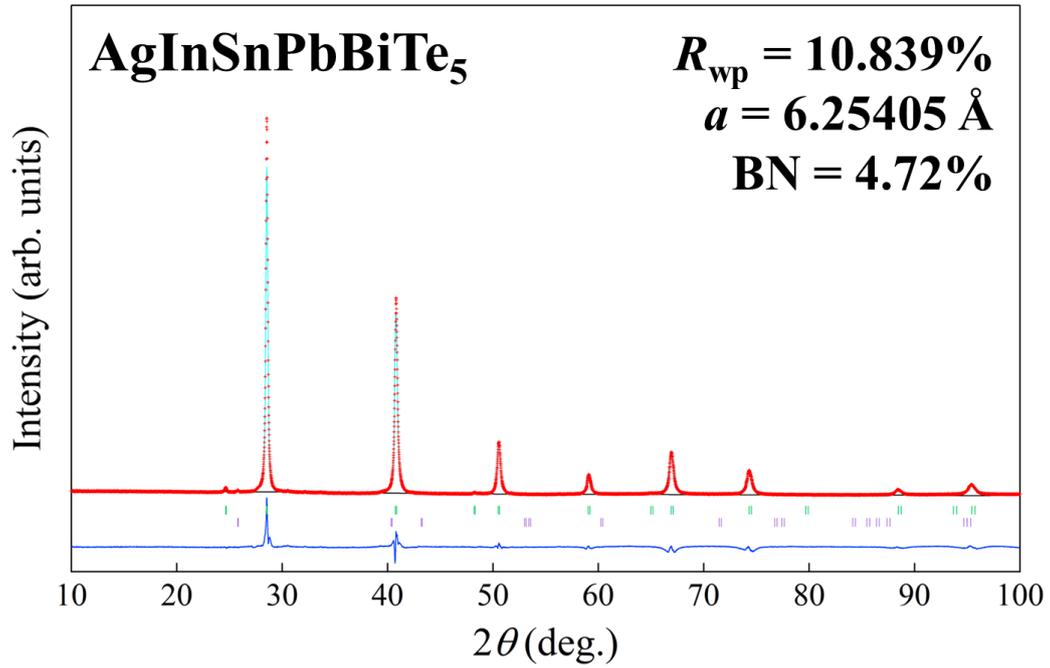

Fig. S1 The result of Rietveld refinement for AgInSnPbBiTe$_5$. Green and purple solid lines represent the peak positions of AgInSnPbBiTe$_5$ and BN, respectively.

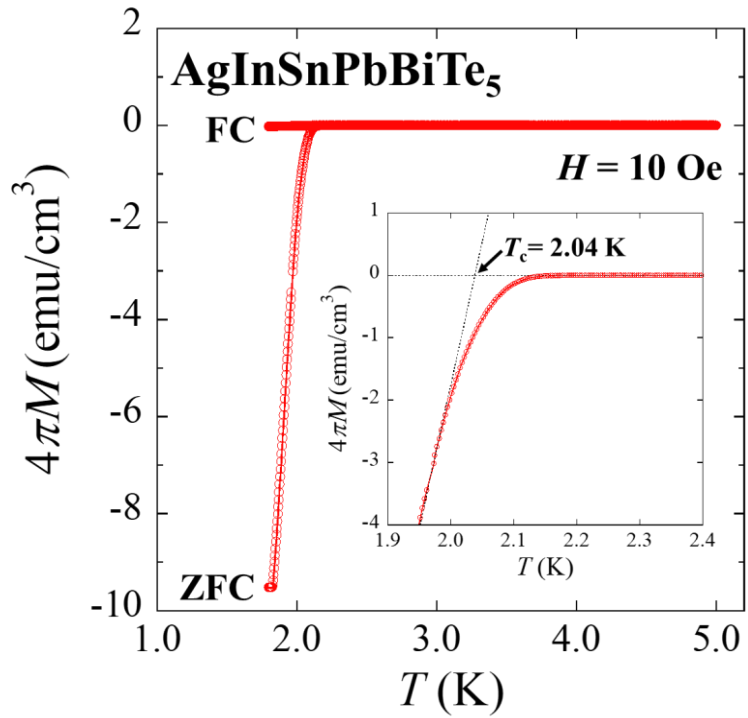

Fig. S2 Temperature dependence of Magnetic susceptibility ($4\pi M$) measured at $H$ = 10 Oe for AgInSnPbBiTe$_5$. Several weeks later from synthesis date, $T_c$ decreased due to the strain release.